\documentclass[%
 reprint,
 amsmath,amssymb,
 aps,
]{revtex4}[two column]

\usepackage{graphicx}
\usepackage{dcolumn,subfig,subcaption,amssymb}
\usepackage{bm}
\usepackage[a4paper,top=26mm,bottom=26mm,left=30mm,right=30mm]{geometry}
\usepackage{tabularx}
\usepackage{amsfonts}
\usepackage{multirow}

\begin{document}

\preprint{APS/123-QED}

\title{Dynamics of Phase Transition in Quark-Gluon Plasma Droplet Formation under Magnetic Field\\}

\author{Agam K. Jha}
 \affiliation{Department of Physics, Kirori Mal college\\ University of Delhi,\\ Delhi-110007, India }


\author{Aviral Srivastava}
\affiliation{
 Delhi Technological University, \\
Shahbad Daulatpur, Main Bawana Road, \\Delhi - 110042, India\\ 
}%



\begin{abstract}Abstract: Pre-existing density of states for a Quark-Gluon Phase, based on Thomas-Fermi and Bethe mode, is expanded by incorporation of new variables. Results from recent study indicate that perturbations in the form of a finite non-zero chemical potential \( \mu \), \( 
\mathcal{\mathbb{B}}\), dynamic thermal masses \( 
\mathcal{\mathbb{M}}\) and of course Temperature \(\mathcal{\mathbb{T}}\) are indeed vital to fully comprehend the formation and dynamics of QGP. Simulations depict an overall increase in the stability of QGP in the paradigm of the statistical model. On the top of Free Energy, Entropy and heat capacity are calculated for the phase transition. The overall qualitative behavior, of entropy or Heat Capacity determines the order of phase transition of the QGP. Investigation of order of phase transition is carried out in this study through Monte-Carlo based differential element, which ensures the inclusion of the randomness of the collisions at the particle colliders.
\end{abstract}

\maketitle


\section{\label{sec:level1}Introduction\protect\\  }
Quantum Chromodynamics (QCD) outlays the varying strength of the strong interaction through the QCD Phase Diagrams. Temperature \(\mathcal{\mathbb{T}}\) and chemical potential \( \mu \) determine the state of in which hadrons exist. QCD is formulized on the SU(3) group renders the calculations much complex. Incorporating new factors in an already complicated system exorbitantly increases computation time. Moreover, the sheer number of dimensions in the QCD Lagrangian and hence the action integral, suggests a statistical approach would be quite precise. Hence, this study extends the existing statistical model\cite{Ramanathan_Mathur_Gupta_Jha_2004}, consistent with URHIC experiments at the Brookhaven National Laboratory (BNL) and CERN. The current prevailing statistical model incorporates temperature and chemical potential\cite{Ramanathan_Jha_Gupta_Singh_2011,PhysRevC.101.045203} as the primary thermodynamic variables, in accordance with the QCD phase diagram. The model successfully predicts the order of phase transition from a confined state to a deconfined state. That being first order, which essentially translates to a discontinuity in the entropy S  against temperature (T). Recent experiments at CERN and BNL suggest the precise description of QGP evolution necessitates the incorporation of new thermodynamic variables, i.e. Magnetic Field. Heavy Au-Au/Pb-Pb ion collisions, need not be head-to-head collisions. \\ \\ Hence it is plausible that the magnetic field is playing a critical role in the formation of the QGP and possibly its physical characteristics. It is important to note that the magnetic field considered here is the external magnetic field \cite{yan2021dynamical,doi:10.1142/S0217751X09047570} present in the environment wherein the QGP droplets form. Another motivation for the incorporation of magnetic fields arises from cosmological and astronomical observations. Primordial magnetic fields of the universe have governed the dynamics of the large-scale magnetic fields that are found in galaxies today. More importantly these primordial were also associated with the phase transitions and inflation in the early universe \cite{PhysRevD.107.014021}. Since QGP phase transitions are a manifestation QCD interactions, magnetic field \cite{Abramchuk:2019lso} must have an effect on the evolution, which interacts through colour charge.  Since QGP prevailed in the very early phase of the universe, it is thus crucial to consider this aspect.\\ \\

\begin{figure}[ht]
    \centering
    \scalebox{0.9}
    {\includegraphics{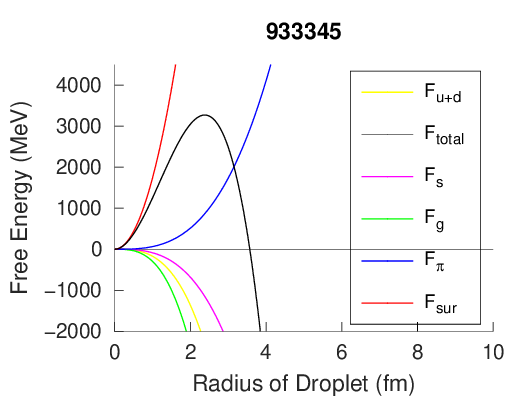}}
    \caption{Variation of Free Energy parametrized by T, \(\mu\) and B.  T = Tc = 175 MeV, \(\mu\)=400 Mev,  \(eB=15m_{\pi}^{2}\)}
    \label{fig:enter-label}
\end{figure}

With an aim to foster a robust and concise model, simple additions have been made to the existing statistical picture. Temperature is a measure of the energy of the system. The collisions which lead to the formation of QGP droplets are relativistic. Further, again from cosmological constraints, the cosmological environment enabled relativistic regimes. Thus, instead of rest masses, dynamical relativistic masses have been used \cite{PESHIER1994235}. Dynamical masses are explicit functions of temperatures. The highlight of this magnetic field study is the parametrization of magnetic field, as an indigenous attempt , strength \cite{PhysRevC.99.035210} on the medium and all the constituents contributing to free energy, which in this case is the \(\pi\)-ionic medium.\\ \\

\section{\label{sec:level1}Discussion\protect\\  }
\subsection{Free Energy}
Through the Thomas-Fermi Model of the atom and the Bethe Model for nucleons, the density of states for quarks and gluons is retained from the previous study cite[1]:\\
\begin{equation} \label{eq:1}
    \rho_{q,g}(k) = \nu(1/\pi)^2\  \Biggl[ [-V_{conf}(k)]^2(\frac{-dV_{conf}}{dk})\biggr]  \\
\end{equation}
Here \(V_{conf}\) is the confining potential \cite{Ramanathan_Mathur_Gupta_Jha_2004, SomorendroSingh:2008fa} for quarks and gluons, given by
\begin{equation}\label{eq:2}
    [V_{conf}]_{q,g} = \frac{1}{2k}(\gamma_{q,g}){g^{2}(k)}.T^{2} - \frac{m^{2}_{o}}{2k}
\end{equation}
All the following parameters are retained: 
\begin{equation}\label{eq:3}
    {g}^{2}(k) = (4/3)(12\pi/27)\biggl[\frac{1}{ln(1+k^{2}/\Lambda^{2})}\biggr] 
\end{equation}\label{QCD coupling constant}
\begin{equation}\label{eq:4}
    k_{min} = (\gamma_{q,g}.N.T^{2}.\Lambda^{2}/2)^{1/4} \\ \vspace{0.1cm}
    where \hspace{0.1cm}\\ \\
    N = (4/3)(12\pi/27)
\end{equation}
\\ \\
\( [-V_{conf}]_{q,g}\), the confining potential constructed from QCD constraints is retained, due to its apt description of the dynamical nature of quarks. While 
 \(\gamma_{q,g}\)  are the hydrodynamic flow parameters, which account for the fluid nature of the plasma. Though a minor altercation is done with the original form of the potential, which is the addition of dynamic temperature-dependent masses. In this statistical model, the Peshier, and thus confining potential is invariant. Thus, the perturbative additions of magnetic field do not affect the Peshier potential constructed from Thomas-Fermi and Bethe models. There exists a simple explanation for this striking result. Magnetic field does no actual work, in the confinement of deconfinement processes, on the system. Peshier potential is a confining potential for the quarks and gluons. The thermal dynamic masses are defined as \cite{PESHIER1994235}\\ 

\begin{equation}\label{eq:5}
    \mathcal{\mathbb{M}}_{d_{q,g}} =  \gamma_{q,g}.T^{2}\biggl[ \frac{1}{log(1+T/T_{c})}\biggr]^{2}\biggl(\frac{16\pi^2}{11 - \frac{2}{3} N_{f_{q,g}}}\biggr)
\end{equation}\\    
where \(T_c\) is the transition temperature, while \(N_{f_{q,g}}\)
is the flavour number which is taken to be 0 for gluons (bosons) and three flavours for quarks (up, down and strange).\cite{PESHIER1994235}. \\

Thus the thermal mass modifies the Peshier potential to yield the following expression

\begin{equation}\label{eq:6}
    [V_{conf}]_{q,g} = \frac{1}{2k}(\gamma_{q,g}){g^{2}(k)}.T^{2} - \frac{\mathcal{\mathbb{M}}_{{d}_{q,g}}}{2k}
\end{equation}
\\
Magnetic field's interplay in the confining potential is disregarded in the context of this simple statistical model. Its interaction with the quark-gluon matter is reflected through the coupling in the partition function and hence ultimately the free energy of the plasma, with a plausible mediation by colour charge. \\ \\
Through these quantities mentioned above, density of states and Peshier potential, the free energy of the QGP constituents is constructed which incorporates three thermodynamical variables as follows: -
\\ \\ \\
\begin{equation}\label{eq:7}
F_{i} =  -T.{g_{g}} \int dk . \rho_{q}(k) . ln\biggl(1 + e^{-{\frac{\bigl(\sqrt{\mathcal{\mathbb{M_{d}}}_{q} + k^2  - (\mathcal{\mathbb{B}})} - \mu_{i}\bigr)}{T}}}\biggr)
\end{equation}
\(F_i  \) denotes the free energy for the constituents quarks of the QGP. Here we consider three flavors of quarks [u,d,s]: up, down and strange. Hence index i goes from 1-3. A similar modified construction can be defined for gluons.

\begin{equation}\label{eq:8}
F_{g} = T.{g_{g}} \int dk .\rho_{g}(k). ln\biggl(1 - e^{-{\frac{\bigl({\sqrt{\mathcal{\mathbb{M_{d}}}_{g} + k^2  -(\mathcal{\mathbb{B})}}}- \mu_{g}\bigr)}{T}}}\biggr)
\end{equation}\\
An additional negative is needed to make the gluon-ic free energy negative.\(\pi - ionic\) medium also exists within the framework of the QGP formation. \( \pi - ions\) manifest as a signature in detection of QGP \cite{Niida2021} at particle colliders. It is highlighted that in this study the medium follows from the FD and BE statistics, as evident from the free energy integrand. Further, unlike the hadrons themselves, the \(\pi\) meson is not modelled to have the dynamic mass. It carries a fix mass of \(\mathcal{M_{\pi}} \sim 138 MeV\) \cite{Fukazawa1991}.A strong reason for this characteristic is the proximal relation of \(\pi\)-ions to chiral symmetry restoration. Their sensitivity to QGP state variables like Temperature \(\mathcal{\mathbb{T}}\) and chemical potential \( \mathcal{\mathbb{T}}\) and magnetic field \(\mathcal{\mathbb{B}}\) holds vital importance under the context of this study.\\ \\
The Free Energy Integral for the \(\pi\)-ionic medium is given by,
\begin{equation}\label{eq:9}
F_{\pi} = - \frac{3T\nu}{2\pi^2} \int dk .k^{2}. ln\biggl(1 + e^{-{\frac{\bigl(\sqrt{\mathcal{{M_{\pi}}}^2 + k^2  + (\mathcal{\mathbb{B}})} - \mu_{i}\bigr)}{T}}}\biggr)
\end{equation}
The Free energy integrals maintain the consistency, which follows from the earlier study. Hence, again statistical distribution functions for Fermi-Dirac (quarks) and Bose-Einstein (Gluons) are being followed. Before addition of a magnetic field, the Free Energy integral follows a positive sign for quarks and a negative sign for gluons. Magnetic Field changes the nature of the integrals and reverses the trend. So, the algebraic signs preceding the integral flip, such an exercise is done on an ad-hoc basis and is vital to ensure the Free energy variation with the QGP droplet radius follows the path towards equilibrium.\\ \\
Finally, the modified surface energy of the droplets is constructed in the presence of magnetic fields. \\ 
\begin{equation}\label{eq:10}
F_{sur} = \frac{\sqrt{2}R^2}{4} \bigl[  \gamma_{f} T^{3} + B*T\bigr] 
\end{equation}\\ \\
In the above equation \ref{eq:10}, \(\gamma_f = \sqrt{(\frac{1}{\gamma_q})^2+(\frac{1}{\gamma_g})^2}\) Dimension of magnetic field is \(MeV^2\) in the natural units system. Coupling of the magnetic field with the system is dependent on the constituents. While for quarks, \(pi\) – ionic medium and surface energy, the field interacts as a square of a coupled quantity {\(\mathcal{\mathbb{B}}\)} and  {\(\mathcal{\mathbb{T}}\)}. On the other hand for bosons in the system, i.e. the gluons, the field interacts without a scaled-down factor of the temperature. Suggesting that coupling between fermions and hadrons is stronger
as compared to gluons.\\ \\

\begin{center}
\begin{tabularx}{0.45\textwidth} { 
  | >{\centering\arraybackslash}X 
  | >{\centering\arraybackslash}X |
}
\hline
Limits in accordance with LHC & Magnetic Field Strength ( \(\frac{\times m_{\pi}^2*15}{0.303})\) \vspace{0.1cm} \\\hline
Minimum & 0.98 \\ \hline
 & 1.107 \\ \hline
 & 1.2 \\ \hline
 & 1.46 \\ \hline
 & 1.71 \\ \hline
 & 1.83 \\ \hline
 & 2.00 \\ \hline
 Maximum & 5.00 \\ \hline

 \end{tabularx}
\end{center}\vspace{0.6cm}
The range of magnetic field is fixed by the LHC exeperiments \cite{PhysRevD.101.034027}, while the supremum is imposed by numerical calculations done within this paradigm. \(\pi\)-ionic free energy and gluon-ic free energy are the the most susceptible to change in magnetic, with the former sets a much more substantial constraint. \\ 
The total free energy is the summation of all the individual constituents without any considerable interaction among them.
\begin{equation}\label{eq:11}
F_total = \biggl(\sum_{i=1}^{3} F_{i}\biggr) + F_{g} + F_{\pi} + F_{sur}
\end{equation}\\
\begin{figure}[ht]
    \centering
    \scalebox{0.9}
    {\includegraphics{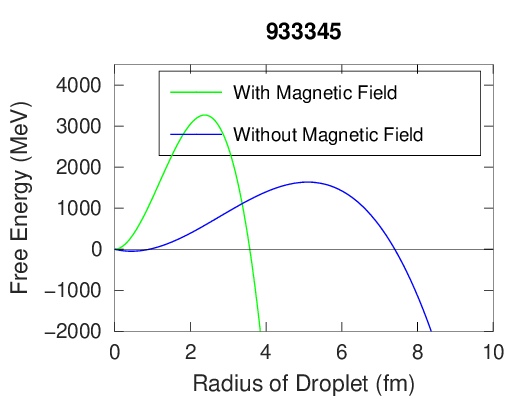}}
    \caption{Comparisson of stability in the absence (blue) and presence (green) of B. With a higher peak indicating a much more stable droplet. T = Tc , \(\mu\)=400 Mev, \(eB=15m_\pi^2\)}
    
\end{figure} \label{fig:enter-label}
The existing free energies of the \(pi\) ionic medium and the surface free energy are parametrized with a magnetic field over the existing thermal dependence. Chemical potential does not play a part in the surface energy. It is modelled in a way which approximates the surface of the droplet in the phase space, which means momentum (k) which is ultimately due to temperature. Implying,  that the surface energy (per unit area) is calculated for a single droplet is same for any droplet of some arbitrary radius and thus does not reflect a summation of the constituents of the system. Their dependence on the magnetic field is modelled which verifies the expected results. In the presence of large magnetic fields produced during the collisions at BNL and CERN, the QGP formation is significantly changed. Magnetic field, restricts the momentum space thus allowing for a greater density of Baryons and thus a higher energy density. Surface Free energy aptly appears like a parallelism to the surface tension of the Quark Gluon Plasma droplet: whose formation is governed by an interplay of QCD interactions and the three dynamical thermodynamic variables in this model. Chemical potential does not appear in the expressions, as explained above. The magnetic field governs the surface structure of each droplet formed through the deconfinement process. \\ \\
Since the droplet is charged (QCD colour charge) and consists of charged constituents (quarks), the magnetic field affects its curvature \cite{Huang2023}. At the equilibrium of the interplay between QCD forces, thermal energy and magnetic field, the droplet assumes a stable radius. 
\(pi\) – ionic free energy.\\\\
Magnetic field does not seem to affect the Peshier potential of the quarks and gluons. Magnetic field is an independent variable. Thus, making the macroscopic states of the fireball system as:\\
i)	Temperature\\ 
ii)	Chemical Potential\\
iii) Magnetic Field\\
\\
\subsection{Entropy, Heat Capacity and Phase Transition}
An intricate perspective towards the formation of the de-confined phase of quarks and gluons is presented through the improved statistical model. To render a wholesome perspective though, a complete statistical and thermodynamic set of state functions is needed. Hence, through the free energy integrals, its derivative are calculated. These derivatives represent different thermodynamic state functions. First derivative of the free energy with temperature is computed, this quantity is the entropy (S) of the system. Subsequently, another derivative with temperature yield the heat capacity \(C_v\) . \\ \\
As the system is transitioning towards equilibrium, entropy is expected to rise with temperature. Which means it will have a positive slope, which is nothing but the graph of the heat capacity. Further, the ratio of Entropy (S) and the heat capacity \(C_v\) define the speed of sound in the plasma phase. Speed of sound is accurately predicted by the QCD calculations. Hence, this quantity serves as a vital test for this extended simple statistical model.
\begin{equation}\label{eq:12}
    \begin{aligned}    
    S_{entropy} = - \frac{\partial F}{\partial T} \\
    C_{heat\hspace{0.1cm}capacity} = T \frac{\partial S}{\partial T} = - T \frac{\partial^2 F}{\partial T^2}
    \end{aligned}
\end{equation}\\ \\
Hence, the speed of sound can be defined as, 
\begin{equation}\label{eq:13}
    C_{s}^2 = S_{entropy}/C_{heat\hspace{0.1cm}capacity}
\end{equation}
In equation \ref{eq:13}, \(C_s^2\) is the square of the speed of sound. As per lattice calculations form Quantum Chromodynamics, this quantity has an supremum, capped at \(1/3\).\\ \\
Entropy and heat capacity also yield information about phase transition. Since formation of Quark-Gluon Plasma involves a change of phase from a confined state of hadrons to a free or de-confined state, the process occurs through a thermodynamic phase transition. In the previous study, with temperature as the only parameter, the order was phase transition was first order. Therefore, a discontinuity in the first derivative of free energy, i.e. entropy was observed. Verification of the nature of phase transition is done again. Mainly to check if its behavior is preserved or undergoes a change. Hence, the plots for entropy, heat capacity and speed of sound are studied to find the order. 
\section{\label{sec:level1}Research Methodology\protect\\  }
\subsection{Literature Review}
The study is reinstated by an overview of the existing statistical model. Followed by a thorough review of new research that has emerged in the field. ALICE experiments at CERN and the RHIC experiments at Brookhaven National Laboratory have fueled the research around this discipline. While the experiments take a long time to run, and an even longer time to collect, analyse annd study the results. Theoretical models, which are developed independently, aid in predicting the result of the experiments. Another vital role played by theoretical models, is in the process of defining constraints. Constraints allow a single run to be effective by dedicating the resources in a dedicated direction which is expected to yield results. Since the experiments reveal new information every time, the review procedure familiarizes with the existing sphere of knowledge around research in QGP. This study presents a novel and indigenous process to comprehend the complex dynamics of QGP formation, so the literature review provided the neccesary outlook which helped in scrutinising existing models and formulating the one employed here.
\subsection{Data for Simulation}
Computations form an integral part in the analysis, first mathematical model was formulated. Which was followed by a simulation through numerical and symbolic functions based computation. Hence, much of the data was generated within the framework of the simulation. As a result, there was no immediate need to collect data from previous studies. Though a variety of existing plots pertaining to the paradigm of the study were referred throughout the study.
\subsection{Qualitative and Quantitative Analysis of results}
Review process prepared a ground work to be built upon. A brief idea of the variation of Free energy in the position space was obtained. However, with the incorporation of every new variable, the system portrayed a different behaviour. The study concludes with an analysis of stability, through a comparison with previous models, analysis of order of phase transition and estimation of speed of sound near the transition temperature.\newpage
\section{\label{sec:level1}Results and Conclusion\protect\\  }
In absence of B thermal motion on account of alignment is small. As \(\mathcal{\mathbb{B}}\) enters the scenario, it competes with thermal energy and counters. So, one expects the degrees of freedom to change, hence the curve shifts to smaller radius values, which is the case. Indeed, this reduction in radius is observed only when the droplet forms i.e. free energy assumes a bell-like behaviour. The quantification of this decrease (in radius) should be proportional to the strength of the field. \(\mathcal{\mathbb{B}}\) thus has an apparent effect on the height of the peak which is proportional to the free energy of the system.\\ \\
\begin{figure}[ht]
    \includegraphics[width=0.5\columnwidth]{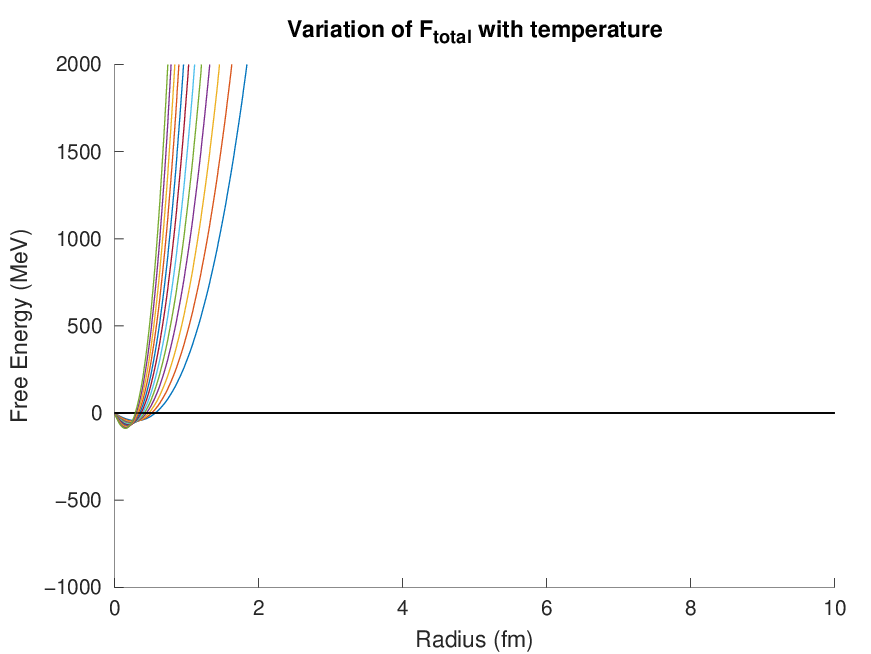}
    \caption{Variation of free energy for different temperatures with \(\mathcal{\mathbb{B}}=0\) and \(\mu = 400 \), with the ratio of flow parameters as 8.}
    \label{mt}
\end{figure} 
As a shift towards dynamic temperature-dependent masses is implemented. It is observed that droplets are not stable (monotonically increasing trends) until chemical potential is added to the distribution function. Which suggests dynamical masses increase the energy content of the system. \\ \\
To reinstate the bell-shaped trends, chemical potential at first and magnetic field reduces the Free energy and enable formation of droplets with a finite peak. A new observation is {\(\mathcal{\mathbb{B}}\)} favours formation of smaller droplets, and hence brings down the mean stable radius of droplets. A high chemical potential is favored by the system ( \(\mu_q=\mu_g=400 \hspace{0.1cm}MeV\) ). The constraint is set by the speed of sound, which is highly sensitive of the chemical potential. A numerical comparison reveals that at \(T = T_c\), average over entire range of magnetic field for a radius of 4 fm shows a 16 times increase in the free energy peak as compared to the basic simple statistical model. \\ \\
\begin{figure}[ht]
    \centering
    \scalebox{0.4}
    {\includegraphics{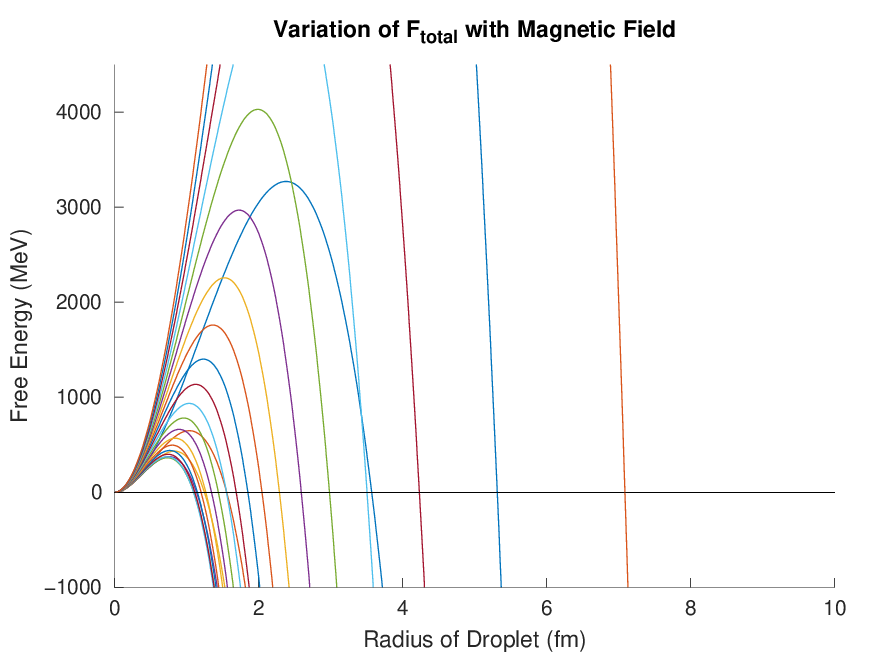}}
    \caption{Variation of Free Energy for various values of B with fixed T and \(\mu\)}
    \label{fig:enter-label}
\end{figure}
\clearpage

\begin{figure}
    \centering
    \includegraphics[width=.3\linewidth]{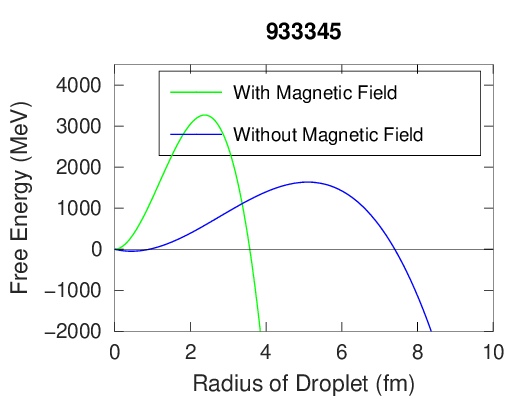}
    \label{fig:sfig2}
\end{figure}

\begin{figure}
    \centering
    \includegraphics[width=.3\linewidth]{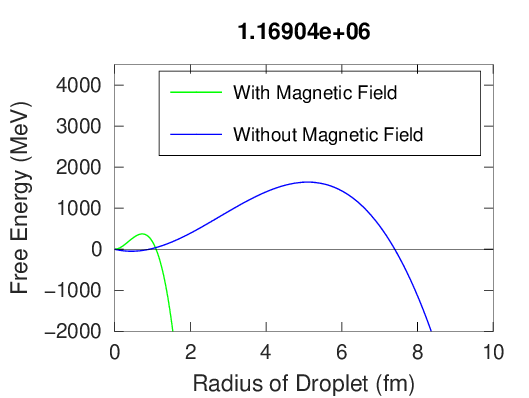}
    \label{fig:sfig2}
\end{figure}
\begin{figure}
    \centering
    \includegraphics[width=.3\linewidth]{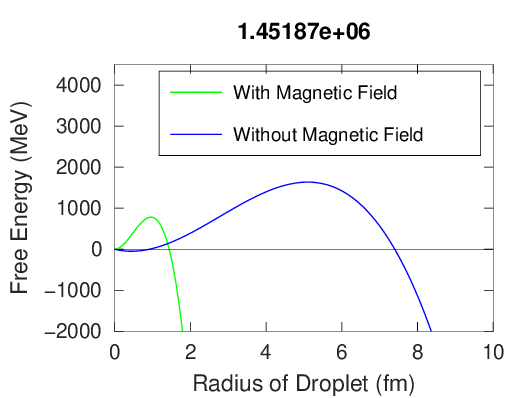}
    \label{fig:sfig2}
\end{figure}
\begin{figure}
    \centering
    \includegraphics[width=.3\linewidth]{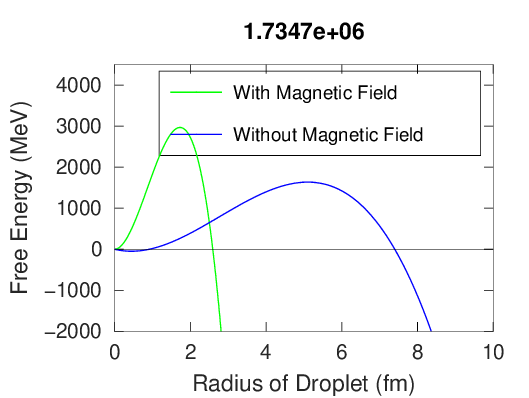}
    \label{fig:sfig2}
\end{figure}
\begin{figure}
    \centering
    \includegraphics[width=.3\linewidth]{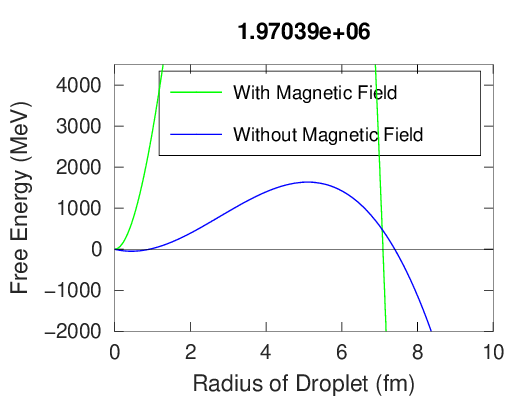}
    \label{fig:sfig2}
\end{figure}

Dependence on the flow parameters has been reduced following the addition of temperature-dependent dynamical masses, chemical potential and magnetic field. The greatest factor affecting this change is chemical potential. Previously, altercation of the flow parameter ratios between quarks and gluons by a difference of 2, the stability of the plasma immensely shifted the trends of the Free Energy curves. The scenario has changed now. Although, the dependence has not vanished entirely, and a factor of \(\gamma_g = ({6,8})\gamma_q\), still give a favorable peak, other variables simply assert dominance. Which means if Magnetic field is high enough, within the appropriate bounds, a feasible phase transition is observed outside the previous stable domain of flow parameters. \\ \\
Finally, the order of Phase transition when all the thermodynamic variables are taken into account is a weakly first-order phase transition. There appears a distinct discontinuity in the trends of entropy near the transition temperature. However, a slight yet continuous deviation in the case of heat capacity is also observed. The ratio of entropy and heat capacity gives the speed of sound ( \(C_s^2\) ). Comparing from the QCD results, which limits the \( C_s^2 \) at 1/3. Under the paradigm of this extended study, it is found that this natural bound is being respected by the trends. Further, the trends obey QCD calculations and computations based on ALICE data. At the effective temperature ( \(222 \hspace{0.1cm}MeV\) ) \cite{Gardim2020}, \(C_s^2{_{T_{eff}}}\) is within \( 16 \%\) margin of error in accordance with the bounds.\\ \\
Concluding this study, it is emphasized that the stability of QGP is greatly enhanced. There are some new observations: dependency on flow parameters and reduced average stable radius which require further study.
\begin{figure}[ht]
    \centering
    {\includegraphics[width=0.45\columnwidth]{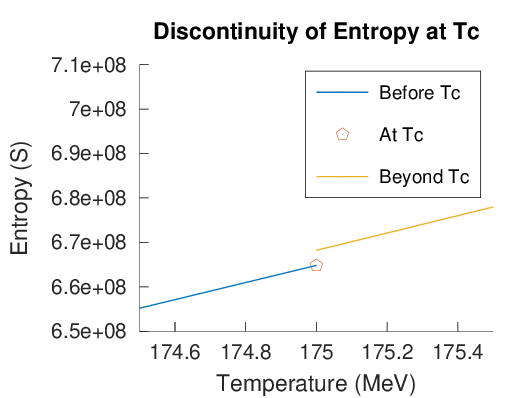}}
    \caption{Discontinuity of Entropy at Transition Temperature}
    \label{fig:enter-label}
\end{figure}
\begin{figure}[ht]
    \centering
    {\includegraphics[width=0.45\columnwidth]{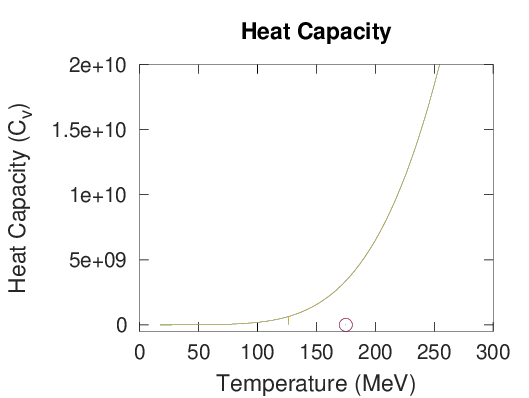}}
    \caption{Continuity of Heat Capacity throughout}
    \label{fig:enter-label}
\end{figure}
\begin{figure}[ht]
    \centering
    {\includegraphics[width=0.45\columnwidth]{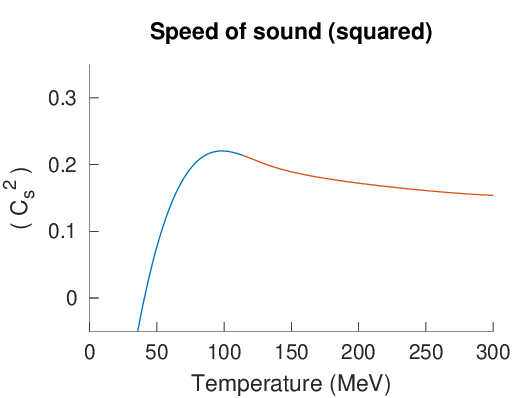}}
    \caption{Asymptotic behavior of Speed of sound(sqaured)}
    \label{fig:enter-label}
\end{figure}\clearpage
\vspace{0.25cm}
\section{\label{sec:level1}Acknowledgements\protect\\  }

We are grateful to Prof. R. Ramanathan (Rtd.), Department of Physics and Astrophysics, University of Delhi, for his vision and unparalleled guidance throughout the course of this study. Further, we would like to thank Delhi Technological University, Dr. M. Jayasimhadri from the Applied Physics Department and Kirori Mal College, University of Delhi for the immense support from these institutions. We have used publicly available open-source software GNU-Ocatve \cite{Octave} for the generation of the plots depicted in this manuscript.
\bibliographystyle{IEEEtran}
\def\bibsection{\section*{References}}
\bibliography{lib.bib}
\end{document}